\documentstyle[prl,aps]{revtex}
\begin{document}
\draft
\preprint{\parbox{1.2in}{UCLA/94/TEP/35 \\ UCONN-94-5}}
\title{Derivative Expansion of the Effective Action and Vacuum Instability \\
for QED in 2+1 Dimensions}
\author{Daniel Cangemi\cite{emaila} and Eric D'Hoker\cite{emailb}}
\address{
Department of Physics,
University of California,
Los Angeles, CA 90024-1547
}
\author{Gerald Dunne\cite{emailc}}
\address{
Department of Physics,
University of Connecticut,
Storrs, CT 06269-3046
}
\date{September 1994}
\maketitle
\begin{abstract}
We investigate the effective action of 2+1 dimensional charged
spin 1/2 fermions and spin 0 bosons in the presence of a $U(1)$ gauge field.
We evaluate terms in an expansion up to second order in derivatives of
the field strength, but exactly in the mass parameter and in the
magnitude of the nonvanishing constant field strength.
We find that in a strong uniform magnetic field background,
space-derivative terms lower the energy, and there arises an instability
toward inhomogeneous magnetic fields.
\end{abstract}
\vfill
\makebox[\textwidth]{UCLA/94/TEP/35 \hfill hep-th/9409113 \hfill UCONN-94-5}

% Use \narrowtext for the submission to PRL
%\narrowtext
\twocolumn

The effective action provides an extremely useful tool for the
investigation of instabilities such as spontaneous symmetry breakdown in
quantum field theory.
These phenomena are frequently driven by low momentum dynamics, so that
a small momentum approximation to the effective action may suffice.
Electrodynamics in 2+1 dimensional space-time (QED$_3$) with massless or
massive
charged spin 1/2 fermions or spin 0 bosons, especially in the presence of a
strong uniform magnetic field, is a field theory model with potentially
many applications in condensed matter and particle physics\cite{condmatt}.
Its dynamics appears intricate and incompletely understood, and may well
reveal exciting new physical phenomena.

In the present paper, we evaluate the effective action for spin 1/2 fermions or
spin 0 bosons of charge $e$ in the presence of a $U(1)$ gauge field in
a derivative expansion\cite{pac}.
We obtain contributions with no derivatives and two derivatives (total) on the
field strength, but our result is exact in the mass of the charged
particle, and is also exact in the magnitude of the field strength.
This calculation -- in fact to all orders in derivatives -- results
entirely from 1-loop effects, and reduces to the evaluation of
functional determinants of spin 1/2 and
spin 0 gauge covariant derivatives\cite{conventions}
\begin{equation}
i \int dx \, {\cal L}_\mp = \pm \log \hbox{Det\ }
\biggl \{ D_\mu D^\mu + m^2 + e \Sigma^{\mu\nu} F_{\mu\nu} \biggr \}
\label{eff_act}
\end{equation}
Here, ${\cal L}_\mp$ are the effective Lagrangians for fermions $(-)$ and
bosons $(+)$, and $D_\mu = \partial_\mu + ie A_\mu$.
For ${\cal L}_-$, $\Sigma^{\mu\nu} = (i/4) [\gamma^\mu, \gamma^\nu]$ produces
the effective action for a 4-component spinor consisting of 2-component spinors
of masses $m$ and $- m$ respectively, whereas for ${\cal L}_+$,
$\Sigma^{\mu\nu} = 0$ produces the effective action for
spin 0 complex scalars, for which the spin coupling term is of course
absent.
The effective action for a single massive 2-component spinor fermion differs
from ${\cal L}_-$
through the inclusion of the parity-violating Chern-Simons term
\begin{equation}
{\cal L}_-^{\text{CS}} = \frac{1}{2} {\cal L}_- + m
\epsilon^{\mu\nu\kappa} A_\mu \partial_\nu A_\kappa
\label{CS}
\end{equation}
as shown in\cite{redlich}.
To obtain a sensible derivative expansion, we assume either that the
derivatives (i.e. the momenta) are small compared to the mass $m$, or
that they are small compared to the background field strength magnitude,
such as the magnitude of a constant magnetic field, or both.
Under these circumstances, we have the expansion (valid for both actions in
(2+1)-dimensional spacetime)
\begin{eqnarray}
{\cal L}^{(2)} & = & p_0
 + e^4 [\partial_\mu F^2 \partial^\mu F^2] \, p_1
 + e^2 [\partial_\mu F_\nu \partial^\mu F^\nu ] \, p_2 \nonumber\\
&& + \, e^2 [\partial_\mu F_\nu \partial^\nu F^\mu ] \, p_3
 + e^6 [F^\mu \partial_\mu F^2 F^\nu \partial_\nu F^2 ] \, p_4
\nonumber\\
&& + \, e^4 [F^\mu \partial_\mu F_\alpha F^\nu \partial_\nu F^\alpha ] \, p_5
 + e^4 [\partial_\mu F^2 F^\nu \partial_\nu F^\mu ] \, p_6 \nonumber\\
\label{pexp}
\end{eqnarray}
up to terms involving at least 3 derivatives.
We find it more convenient to use the ``dual'' field
strength
$F^\mu = \frac{1}{2} \epsilon^{\mu\nu\kappa} F_{\nu\kappa}$, and $F^2 = F_\mu
F^\mu =
\frac{1}{2}F_{\mu\nu} F^{\mu\nu}$, with $B = - F_0$,
$E^i = - \epsilon^{0ij} F_j$ the usual magnetic and electric fields.
We have made use of the fact that $F^\mu$ satisfies the Bianchi
identities $\partial_\mu F^\mu = i (\partial E - \bar\partial \bar E) - \dot B
= 0$ to restrict possible contributions in~(\ref{pexp}).
The coefficients $p_i$ are functions of $e^2 F^2 = e^2 (B^2 - {\vec E}^2)$, and
of
the mass parameter $m$.
They are Lorentz and gauge invariant and even under $F^\mu\rightarrow -
F^\mu$.
Notice that in (3) we have not retained terms that are odd under $F^\mu
\rightarrow - F^\mu$, even though they would be allowed by Lorentz
and gauge invariance (for example $\epsilon^{\mu\nu\kappa} \partial_\mu
F^\alpha \partial_\alpha F_\nu F_\kappa$ and $\epsilon^{\mu\nu\kappa}
\partial_\mu F^2  F^\alpha \partial_\alpha F_\nu F_\kappa$).
Such terms have vanishing contribution in view of charge conjugation
symmetry of~(\ref{eff_act}), a property usually referred to as Furry's theorem.

The above expansion may easily be rewritten in terms of $\vec E$
and $B$, which may be particularly useful when investigating dynamics
around large constant magnetic fields.  We can, in fact,
exploit the global Lorentz invariance of the effective action to express the
effective Lagrangian (\ref{pexp}) in a Lorentz frame in which the
constant part of $\vec{E}$
vanishes\cite{complex}
\begin{eqnarray}
{\cal L}^{(2)} &=& q_0
 + e^2 [\bar\partial B\partial B ] \, q_1
 + e^2 [\dot{\bar E}  {\dot E} ] \, q_2 \nonumber\\
&& + \, e^2 [\bar\partial {\bar E} \partial E ] \, q_3
 + e^2 [\bar\partial E \partial \bar E ] \, q_4 \nonumber \\
&& + \frac{1}{2} e^2
 [\partial E \partial E + \bar\partial \bar E \bar\partial \bar E ] \, q_5
 - \frac{i}{2} e^2 [\partial B \dot E - \bar\partial B
 \dot{\bar E} ] \, q_6 \nonumber\\
\label{qexp}
\end{eqnarray}
where the $q_i$ are functions of $e B$ with the renormalization condition
$q_0(0) = 0$.
The functions $p_i$ and $q_i$ are algebraically related as
follows:
\begin{eqnarray}
p_0 &=&    q_0 \nonumber\\
p_1 &=& - (q_1 + q_3 + q_4 + q_5) / (16 e^2 F^2) \nonumber\\
p_2 &=&   (q_3 + q_4 + q_5) / 4 \nonumber\\
p_3 &=& - (q_3 - q_4 + q_5) / 4 \label{pq}\\
p_4 &=&   (q_1 + 4 q_2 + q_3 + q_4 - q_5 - 2 q_6) / (4 e^2 F^2)^2 \nonumber\\
p_5 &=& - (4 q_2 + q_3 + q_4 + q_5) / (4 e^2 F^2) \nonumber\\
p_6 &=&   (q_3 - q_4 + q_5 + q_6) / (4 e^2 F^2) \nonumber
\end{eqnarray}

In the remainder of the paper, we shall determine the functions $p_i$
and $q_i$ explicitly, using the Schwinger proper time method\cite{schwinger},
reformulated in terms of quantum mechanical path integrals over closed
loops $y^\mu(\tau+T) = y^\mu(\tau)$ and for fermions also with
additional one-dimensional Grassmann
variables\cite{boundary}
$\psi^\mu(\tau+T) = -\psi^\mu(\tau)$.
This method is particularly convenient to use in the background of
constant ${\vec E}$ and $B$ fields, where ordinary Feynman diagram
techniques are cumbersome\cite{redlich,strassler}.
The effective Lagrangian is given in dimensional regularization around $d = 3$
by the following
expectation value\cite{euclidean}
\begin{equation}
{\cal L}_\mp =  C_\mp \int_0^\infty \frac{dT}{T}
(2\pi {\cal E} T)^{-d/2} e^{-m^2 {\cal E} T/2} \Bigl\langle e^{ \int_0^T d\tau
L_\mp^{\text{I}} } \Bigr\rangle_{L_\mp^{\text{free}}}
\label{prop_time}
\end{equation}
with $C_- = - 1$ for a 2-component spinor, $C_- = - 2$ for a 4-component spinor
and $C_+ = 1$ for a complex scalar.
Here the free and interacting Lagrangians are given by
\begin{eqnarray}
\lefteqn{
L_+^{\text{free}} = \frac{1}{2 {\cal E}}\dot y_\mu \dot y^\mu} \hspace{1.4in}
&&
L_+^{\text{I}} = - ie A_\mu (x_o+y) \dot y^\mu \nonumber\\
\lefteqn{
L_-^{\text{free}} = L_+^{\text{free}} + \frac{1}{2} \psi_\mu {\dot \psi}^\mu}
\hspace{1.4in}&&
L_-^{\text{I}} = L_+^{\text{I}} + \frac{ie}{2} {\cal E} \psi^\mu F_{\mu\nu}
(x_o+y) \psi^\nu \nonumber\\
\label{lagrangian}
\end{eqnarray}

The vacuum expectation value in~(\ref{prop_time}) is taken with respect to the
free
Lagrangian, $x_o$ is the average position of the closed loop at which
${\cal L}_\mp$ is evaluated and $\int\nolimits_0^\tau d\tau~y(\tau) = 0$.
To evaluate ${\cal L}_\mp$ in a derivative expansion, we expand
$L_\mp^{\text{I}}$ in derivatives of $A$ and $F$.
In the Fock-Schwinger gauge, we have
\begin{eqnarray}
A_\mu(x_o+y)  = \frac{1}{2} y^\rho F_{\rho\mu} (x_o) + \frac{1}{3} y^\sigma
y^\rho
\partial_\sigma F_{\rho\mu} (x_o) \nonumber\\
 + \frac{1}{8} y^\omega y^\sigma y^\rho \partial_\omega \partial_\sigma
F_{\rho\mu} (x_o) + \cdots
\label{FS_gauge}
\end{eqnarray}
We need to retain second order derivatives in this expansion to linear
order, since by integration by part in $x_o$, they yield terms bilinear
in single derivatives of $F$.

As is well-known, the constant $F_{\mu\nu}$ problem is quad\-rat\-ic and may
be solved completely\cite{schwinger,blau}.
The derivative expansion we are interested in is thus a perturbation
around constant $F_{\mu\nu}$.
We denote by $L^n$ for $n\geq 0$, the contribution to the
interaction Lagrangian $L^{\text{I}}$ resulting from the $n$-th expansion term
in~(\ref{FS_gauge}), containing $n$ derivatives on $F$.
It is convenient to rearrange the expectation value of~(\ref{prop_time})
as
\begin{eqnarray}
\lefteqn{
\Bigl\langle\exp \int_0^T d\tau L^{\text{I}} \Bigr\rangle_{L^{\text{free}}}
= } \nonumber\\
&& \Bigl\langle \exp \int_0^T d\tau L^0 \Bigr\rangle_{L^{\text{free}}}
\Bigl\langle \exp \int_0^T d\tau (L^{\text{I}} - L^0)
\Bigr\rangle_{L^{\text{free}}+L^0} \nonumber\\
\label{const}
\end{eqnarray}
To the order we are interested in, the exponential in the second factor
may be expanded, and we get
\begin{eqnarray}
\lefteqn{
\Bigl\langle\exp \int_0^T d\tau (L^{\text{I}} -
L^0) \Bigr\rangle_{L^{\text{free}} + L^0} =
 1 + \int_0^T d\tau \bigl\langle L^2 \bigr\rangle_{ L^{\text{free}}
+L^0}} \hspace{.5in} \nonumber\\
&& + \frac{1}{2} \int_0^T d\tau \int_0^T d\tau'
\bigl\langle L^1(\tau) L^1(\tau') \bigr\rangle_{L^{\text{free}} + L^0}
\label{der_exp}
\end{eqnarray}
The first factor in~(\ref{const}) is just the constant electromagnetic field
problem, and is easily evaluated
\begin{eqnarray}
\lefteqn{
\Bigl\langle\exp \int_0^T d\tau~L_\mp^0 \Bigr\rangle_{L_\mp^{\text{free}}} =
} \hspace{.5in} \nonumber\\
&& \left\{ \begin{array}{ll} (bT/2) \coth (bT/2) & \mbox{for $(-)$ fermions} \\
                          {(bT/2) / \sinh (bT/2)} & \mbox{for $(+)$ bosons}
\end{array} \right.
\end{eqnarray}
We shall henceforth use the abbreviation $b = e (F^2)^{1/2}$.
To obtain the correction from derivative terms of $F$ in~(\ref{der_exp}), we
need
$y^\mu$ and $\psi^\mu$ propagators, in the presence of constant
$F_{\mu\nu}$ fields.
They are
\begin{eqnarray}
\bigl\langle y^\mu y^\nu \bigr\rangle & = & (\eta^{\mu\nu} - \hat F^\mu \hat
F^\nu) G_0 + \hat F^\mu \hat F^\nu G_1 + i~\epsilon^{\mu\nu\kappa}
\hat F_\kappa G_2 \nonumber\\
\bigl\langle \psi^\mu \psi^\nu \bigr\rangle & = &
(\eta^{\mu\nu} - \hat F^\mu \hat F^\nu)S_0 + \hat F^\mu\hat F^\nu S_1 +
i~\epsilon^{\mu\nu\kappa }\hat F_\kappa S_2 \nonumber\\
\end{eqnarray}
where $\hat F^\mu = F^\mu (F^2)^{-1/2}$.
The scalar functions $G$ and $S$ are given as functions of $\tau =
\tau_1-\tau_2$
\begin{eqnarray}
G_0(\tau) & = & - \frac{1}{2 b} \frac{\cosh(b |\tau|- b T/2)}{\sinh (b T/2)} +
\frac{1}{b^2 T}\nonumber\\
G_1(\tau) & = &  - \frac{1}{2T} |\tau|^2 + \frac{1}{2}|\tau| -
\frac{T}{12}\nonumber\\
S_0(\tau) & = & - \frac{1}{2} \epsilon(\tau) \frac{\cosh(b |\tau| - b
T/2)}{\cosh
(b T/2)} \label{green}\\
S_1(\tau) & = & - \frac{1}{2} \epsilon(\tau)\nonumber
\end{eqnarray}
Here $\epsilon(\tau) = +(-)$ for $\tau > 0 (< 0)$, and the functions
$G_2$ and $S_2$ will not be needed explicitly, except for the fact that
$\dot G_2 = - b G_0, \dot S_2 = - b S_0$.

{}From the above formalism, we obtain the following results for the
functions $q_i$ in~(\ref{qexp}):
\begin{equation}
q_i = \left( \frac{1}{4\pi e B} \right)^{3/2}
 \int _0^\infty ds\, e^{- \frac{m^2}{e B}s}\, s^{-3/2} \, f_i(s)
\label{qi}
\end{equation}
In the fermion case, they are expressed in terms of the function $\ell_-(s)
\equiv s \coth s$:
\begin{mathletters}
\label{fi}
\begin{eqnarray}
f_0 &=& - (e B)^3 \, \frac{1}{s} (\ell_- - 1) \nonumber\\
f_1 &=& - \frac{s}{2}\, \ell_-^{\prime\prime\prime} \nonumber\\
f_2 &=& - \frac{3}{4 s}\, \ell_-^{\prime} + \frac{1}{2}
\nonumber\\
f_3 &=& - \frac{1}{6} (s^2 - 3) \ell_-^{\prime\prime} - \frac{1}{3} (s^2 + 3)
\frac{1}{s} \ell_-^{\prime} + \frac{1}{3} \ell_- \nonumber\\
f_4 &=& \frac{1}{8} \ell_-^{\prime\prime} + \frac{1}{2 s} \ell_-^{\prime} -
\frac{1}{6} \ell_- - \frac{1}{4} \nonumber\\
f_5 &=& - \frac{1}{12} (3 - 2 s^2) \ell_-^{\prime\prime} - \frac{1}{6} (s^2 -
3) \frac{1}{s} \ell_-^{\prime} + \frac{1}{3} \ell_- - \frac{1}{2} \nonumber\\
f_6 &=& - \frac{1}{4} \ell_-^{\prime\prime} - \frac{1}{6} (s^2 + 3) \frac{1}{s}
\ell_-^{\prime} + \frac{1}{2}
\end{eqnarray}
whereas in the boson case, they are written in terms of $\ell_+ \equiv s /
\sinh s$:
\begin{eqnarray}
f_0 &=& (e B)^3 \, \frac{1}{s} (\ell_+ - 1) \nonumber\\
f_1 &=& \frac{s}{2} \ell_+^{\prime\prime\prime} + \frac{s}{2} \ell_+'
\nonumber\\
f_2 &=& \frac{3}{4 s} \ell_+^{\prime} + \frac{1}{4} \ell_+
\nonumber\\
f_3 &=& \frac{1}{6} (s^2 - 3) \ell_+^{\prime\prime} - \frac{1}{6} (s^2 - 6)
\frac{1}{s} \ell_+^{\prime} + \frac{1}{6} \ell_+ \nonumber\\
f_4 &=& - \frac{1}{8} \ell_+^{\prime\prime} - \frac{1}{2 s} \ell_+^{\prime} -
\frac{5}{24} \ell_+ \nonumber\\
f_5 &=& \frac{1}{12} (3 - 2 s^2) \ell_+^{\prime\prime} + \frac{1}{6} (s^2 - 3)
\frac{1}{s} \ell_+^{\prime} - \frac{1}{12} \ell_+ \nonumber\\
f_6 &=& \frac{1}{4} \ell_+^{\prime\prime} + \frac{1}{6} (s^2 + 3) \frac{1}{s}
\ell_+^{\prime} + \frac{1}{4} \ell_+
\end{eqnarray}
\end{mathletters}
The integrals in (\ref{qi}) can all be expressed explicitly in terms of
generalized Riemann
zeta-functions\cite{zeta}.

It is instructive to consider two important physical limits. In the
massless limit, $m=0$, we have
\begin{equation}
\left[{\cal L}_\mp^{(2)}\right]_{m=0} = - \frac{(e B)^{3/2}}{\sqrt{2} (2\pi)^2}
\left( \alpha_\mp + \beta_\mp \, \frac{\bar{\partial} B\partial B}{e B^3}
\right)
\label{zeromass}
\end{equation}
with $\alpha_- = \zeta(3/2) \approx 2.6$, $\beta_- = - (15/16\pi) \, \zeta(5/2)
\approx - 0.4$ and $\alpha_+ = (1 - 1/\sqrt{2}) \, \zeta(3/2) \approx 0.8$,
$\beta_+ = - (\sqrt{2} - 1) $ $ \pi/4 \, \zeta(1/2) - (1 - 1/(2\sqrt{2}))
(15/16\pi) \, \zeta(5/2) \approx 0.2$.
Notice that ${\cal L}^{(2)}$ diverges in the $B\to 0$ limit for
massless particles.
Rather, the small $B$ limit should be taken relative to the
scale set by the fermion or boson mass. Thus, one should expand the effective
Lagrangian
in terms of the ratio of the cyclotron energy scale $e B/m$ and the rest mass
energy scale $m$:
\begin{mathletters}
\label{smallb}
\begin{eqnarray}
{\cal L}_-^{(2)} &=& - \frac{m^3}{24\pi} \left( \case{e B}{m^2} \right)^{2} +
\frac{\bar{\partial} B\partial B}{e B^3} \frac{m^3}{60\pi} \left( \case{e
B}{m^2} \right)^3 + \cdots \\
{\cal L}_+^{(2)} &=&  - \frac{m^3}{48\pi} \left( \case{e B}{m^2} \right)^{2} +
\frac{\bar{\partial} B\partial B}{e B^3} \frac{m^3}{240\pi} \left( \case{e
B}{m^2} \right)^3 + \cdots
\end{eqnarray}
\end{mathletters}

The most immediate physical consequence to be drawn from this work
concerns the stability of a state in which the background electric and
magnetic fields are non-zero.
In four dimensional QED, a uniform electric background field, produces
an instability which leads to the spontaneous creation of
electron-positron pairs.
In the present case of three dimensional QED, the same instability
exists for electric fields.
In addition however, there now also arises an instability related purely
to magnetic fields.

For both bosons and fermions, the presence of a uniform magnetic
fields increases the energy of the state for all values of the mass, as
can be seen directly from~(\ref{qi}) and~(\ref{fi}) ($f_0(s)$ is a negative
function).
The presence of inhomogeneities in the magnetic field on the other hand
may lower the energy.
For large mass ($m \gg (e B)^{1/2}$), the leading derivative
terms in the effective action (\ref{smallb}) have a positive coefficient and
lower
the energy as soon as inhomogeneities are introduced.
For bosons, this phenomenon disappears when the mass falls below a
certain critical value $m \approx 0.9 (e B)^{1/2}$.
For fermions however, the sign of the derivative term does not change
with mass and inhomogeneities in the magnetic field always lower the
energy.
Our conclusions are of course limited to the approximation in which the
derivatives on the magnetic field are much smaller than either $(eB)^{1/2}$
or $m$.

Physically, the magnetic field itself is dynamical and we
briefly discuss how the above conclusions are modified.
Dynamics in QED$_3$ is usually introduced through the
Maxwell Lagrangian $(\vec E^2-B^2)/2$ or through the Abelian Chern Simons term,
or both.
If only the Maxwell term is added, the conclusions of the preceding
paragraph are modified.
Indeed, we may then focus on perturbations around constant magnetic
field that are time-independent, so that no fluctuations in the electric
field arise to this order.
Again, these perturbations will be inhomogeneous, they will lower the
energy and create an instability.
When a Chern-Simons term is present, the electric field couples directly
to the magnetic field and all terms in the effective action (\ref{qexp}) should
be retained in the analysis.
It is possible that under these circumstances, the constant magnetic
background state is stabilized, but we have not completed the
investigation of this effect.

The above conclusions may be relevant to some recent proposal concerning
the stability of the $B=0$ state in QED$_3$ and the
possible associated breaking of Lorentz invariance.
In\cite{hosotani}, a version of QED$_3$ is proposed with chiral
fermions and a bare Chern-Simons term, arranged precisely in such a way
as to cancel the induced Chern-Simons term of (\ref{CS}).
It is argued in\cite{hosotani} that dynamical fluctuations in the
electro-magnetic
fields are responsible for an instability of the $B = 0$ state and that
a state with non-zero uniform magnetic field is the correct ground
state.

We reconsider these assertions in light of the above results. First of all, our
use of the effective action has the advantage that Lorentz
invariance is preserved at all stages of the calculation. Then, if
indeed a ground state were to arise with non-zero and uniform vacuum
expectation value for the magnetic field, we can use the above
analysis to study the dynamics of small fluctuations around the proposed
state.
As shown above, time-independent fluctuations produce an instability
towards inhomogeneities in the magnetic field.
It appears that the uniform magnetic field state is not a stable one,
but restructures itself in an inhomogeneous pattern with lower energy.
Thus, the conclusions of\cite{hosotani}, based upon the assumption that the
ground
state is supported by a uniform magnetic field appear to deserve further
investigation.

The present analysis itself may however shed light on the nature of the true
ground state of QED$_3$.
For example, our analysis could be used to extend the results of
Ref.\cite{miransky}, concerning dynamical flavour symmetry breaking in QED$_3$
by a magnetic field, to the case where inhomogeneities are present. Furthemore,
from some points of view, this theory is similar to four-dimensional QCD.
It was shown in\cite{polyakov} that (compact) QED$_3$ confines
electric charges with a linear potential, just as in QCD.
This confinement comes about because
instantons disorder magnetic and electric fields. From this analogy, one
may reasonably conjecture that the true ground state of three-dimensional QED
is more like the QCD ground state with disordered magnetic fields, than like
an
ordered uniform magnetic field. Our calculations indeed show an instability
of the uniform magnetic field state towards a more disordered state
with inhomogeneous magnetic fields.

We thank Z.~Bern, Y.~Hosotani, S.~Kivelson and V.~Miransky for
useful discussions. One of us (E.D'H.) thanks the Aspen Center for Theoretical
Physics for its hospitality.
This work is supported in part by the NSF under contract
PHY-92-18990, by the Swiss National Science Foundation and
by the DOE under grant DE-FG02-92ER40716.00.


\begin{references}
\bibitem{conventions} Minkowski spacetime has signature
$(1, - 1, - 1)$ and the
completely antisymmetric tensor in (2+1)-dimensions is defined by
$\epsilon^{012} = 1$.
\bibitem{complex} We use the convenient complex notation $E\equiv(E^1 + i E^2)$
and $\partial \equiv (\partial_1 - i \partial_2) / 2$.
\bibitem{boundary} The anti-periodicity of $\psi^\mu$ can be established
directly by using the coherent state formulation of the fermion path integral,
or may be seen from the low energy limit of superstring theory, where spacetime
fermions in the Ramond sector obey antiperiodic boundary conditions on the
disk.
\bibitem{euclidean} The path integral formulation is actually carried in
Euclidean space (${\cal E} = 1$) and then analytically continued to Minkowski
(${\cal E} \to i, x^0 \to i x^0, \psi^0 \to i \psi^0, A_0 \to i A_0, A_i \to -
A_i$) to give expressions (\ref{prop_time}) and (\ref{lagrangian}). It is
possible to deform the integral over $T$ on negative imaginary values; thus we
henceforth set ${\cal E} = 1$.
\end{references}
\end{document}